\documentstyle[12pt]{article}
\begin{document}

\centerline{\bf Analysis of Pion Photoproduction} 
\centerline{\bf over the Delta Resonance Region }
\vskip .3cm
\centerline{R.A. Arndt, I.I. Strakovsky, and R.L. Workman } 
\centerline{Department of Physics, Virginia Tech, Blacksburg, VA 24061 }

\begin{abstract}
A number of recent multipole analyses from Mainz, RPI, BNL and Virginia Tech 
(VPI) have focused on the first resonance region. 
One goal common to these studies 
was an improved set of $\Delta$(1232) photo-decay amplitudes. There has been
considerable debate over the differing results found for A$_{1/2}$, A$_{3/2}$ 
and the E2/M1 ratio. We show that a much more consistent set of values is 
possible if differences in the database are considered.  
\end{abstract}
\vskip .5cm
\centerline{\bf RECENT MULTIPOLE ANALYSES}
\vskip .3cm
The importance of the $\Delta$(1232) photo-decay amplitudes in constraining
quark models is well known. While those constructing models have struggled
to reproduce the `observed' scale and ratio of the photo-decay amplitudes,
a similarly vigorous program has been launched by both the experimentalists
and the phenomenologists. Very recently, the Mainz\cite{beck} and 
BNL\cite{bnl} groups have released differential cross section and   
beam-asymmetry ( $\Sigma$ ) data, along with their own multipole analyses. 
These new data have also been analyzed by the VPI\cite{vpi}, RPI\cite{rpi},
and Mainz theory\cite{mainz} groups. Having so many groups working on
essentially the same problem brings to mind a quote (attributed to 
Lovelace) which can be found in H\"ohler's book\cite{hoehler} 
on $\pi N$ scattering.
\vskip .3cm
\centerline{ ``To have five groups trying to cut each other's }
\centerline{  throats is more efficient than random searching"}
\vskip .3cm
While there may be some truth to this,
the intended purpose of the present paper is to search for unity rather than
diversity in the various approaches. This is actually possible if the
database is critically examined. However, before doing so, we should
outline how the various analyses differ.

The VPI analyses\cite{vpi} are either energy-dependent (based on a K-matrix
approach) or energy-independent. In the energy-independent analyses, data in
a narrow window of energy are analyzed together. Here multipoles take their
phases from the energy-dependent fit; the moduli are allowed to vary. 
Results for the photo-decay amplitudes (A$_{1/2}$ and A$_{3/2}$) agree with
the Mainz and RPI results, within uncertainties, but are below the BNL 
value for A$_{3/2}$. The E2/M1 ratio has generally been lower (in magnitude)
than those found in more recent analyses. The reason for this is given below.

The analysis made by the Mainz theory group\cite{mainz} employs fixed-t
dispersion relations for the multipoles (as opposed to fixed-t relations
for the invariant amplitudes). Details are given in Tiator's contribution
to this symposium.

The BNL analysis\cite{bnl} fits both pion photoproduction and Compton 
scattering data (linked by common systematic errors). 
The form used in describing the 
photoproduction multipoles is similar to that used in the VPI analysis.
The multipoles are further constrained via dispersion relations for Compton
scattering. As mentioned above, the BNL value for A$_{3/2}$ is somewhat
larger (in magnitude) than those reported in other recent analyses.

The RPI analysis\cite{rpi} differs from those already described in that an
effective Lagrangian is used and only five parameters\cite{rick} 
are fitted to the data. The value found for the E2/M1 ratio is slightly
larger (in magnitude) than that found in the Mainz fits over a similar
database. 

The Mainz experimental group\cite{beck} has determined the E2/M1 ratio from
a polynomial fit ( in $\cos (\theta )$ ) to the beam-polarized cross sections.
One particular ratio of the polynomial coefficients corresponds to the E2/M1
ratio, if other terms and the interference of other multipoles can be shown
to be small. No values for A$_{1/2}$ and A$_{3/2}$ are extracted. 

\vskip .5cm
\centerline{\bf EFFECT OF CHANGES IN THE FITTED DATABASE}
\vskip .3cm
Much of the confusion surrounding the E2/M1 ratio can be traced back to
the use of different databases in the various fits. Not surprisingly, 
the Mainz\cite{mainz} and BNL\cite{bnl} fits were based largely on data 
produced at their own facilities. The RPI fit\cite{rpi} 
was similarly restricted to include mainly the Mainz data\cite{beck} over the 
resonance. The initial VPI fits\cite{vpi} were unique in that they included
the entire database. Changes between the previous and most recent VPI 
amplitudes were due mainly to the addition of Mainz differential cross 
section, $\Sigma$, and total cross section data\cite{beck,mac}.

More restrictive fits have now been made in order to
track down the source of the differing results. Our first and most severe 
test was a fit which rejected all pre-1980 cross section data. 
By considering how the fit to individual data had changed, 
we managed to narrow the cut down to just two sources
of $\pi^0 p$ differential cross sections.  
These are two Bonn sets\cite{gz74} 
which are essentially consistent with the more 
recent Mainz measurements. Our analysis of this modified data set (F500) 
resulted in $\Delta$(1232) parameters which are in close agreement with the
Mainz values\cite{mainz}. These are displayed in Table~I. 
This observation is also supported by a remark 
made by the BNL group\cite{bnl}. 
In that work, as a test, the BNL cross sections were removed and 
replaced by the Bonn cross sections. The result was
an E2/M1 ratio of $-1.3$\%. Preliminary fits by Davidson\cite{rick} appear to
show a qualitatively similar behavior. The addition of more cross section
data tends to lower the E2/M1 ratio.  

The above result seems to contradict the often repeated remark that:
``the Mainz and Bonn cross sections agree''. A more appropriate statement 
would be that: ``the Mainz and Bonn cross sections agree over the angular
range where they can be compared''. At the resonance energy, the Bonn 
cross sections\cite{gz74} range from 10$^\circ$ to 160$^\circ$ while the
Mainz cross sections\cite{beck} go from 75$^\circ$ to 125$^\circ$. The Bonn
data (for neutral pion photoproduction) are not well described by analyses 
which have excluded them.  

Finally, we note that the BNL E2/M1 ratio\cite{bnl} is slightly larger
in magnitude than the Mainz value. Understanding this difference is 
complicated by the fact that BNL carried out a joint analysis of their 
pion photoproduction and Compton scattering data. We can, however, ask 
whether the new BNL $\Sigma$ data for $\pi^0 p$ and $\pi^+ n$ are mainly 
responsible for the difference. In order to address this question, we  
included the new BNL $\Sigma$ data in a low-energy fit. The result 
was a solution (B500) with essentially the same E2/M1 ratio. 
We should also note that the Mainz analysis\cite{mainz} appears to fit this 
$\pi^0 p$ $\Sigma$ data\cite{bnl} quite well (though it was not included).
Thus, it would seem that differences in the cross section and/or the 
influence of the Compton scattering data are mainly responsible for the 
slightly higher BNL ratio.

\begin{table}[tbh]
\caption{Results for photo-decay parameters from our most recent 
publication (W500) (using the full SAID database), 
from a fit (F500) using a
restricted SAID database (see text), a similar fit (B500) which
includes recent BNL $\Sigma$ data, 
and the Mainz analysis.}
\label{tbl1}
\begin{center}
\begin{tabular}{|c|c|c|c|c|}
\hline
Fit      &  A$_{1/2}$   &  A$_{3/2}$  & E2/M1  &  E2/M1(pole) \\
\hline         
         &  ($10^{-3}$)GeV$^{-1/2}$  &  ($10^{-3}$)GeV$^{-1/2}$ & & \\
\hline
W500     &  $-135(5)$   &  $-250(8)$  &  $-1.5(5)$\%  
         &  $-0.034(5)-0.055(5)i$ \\
F500     &  $-130(4)$   &  $-250(6)$  &  $-2.6$\%  &  $-0.033-0.043i$ \\
B500     &  $-129(2)$   &  $-248(3)$  &  $-2.5$\%  &  $-0.032-0.049i$ \\
Mainz    &  $-129(2)$   &  $-247(4)$  &  $-2.4$\%  &  $-0.035-0.046i$ \\
\hline
\end{tabular}
\end{center}
\end{table}

The status of the photo-decay amplitudes is much clearer. 
The results for A$_{1/2}$ and A$_{3/2}$ from VPI/RPI/Mainz 
agree within uncertainties. The only significant problem 
is posed by the BNL result which is probably a reflection of their 
larger cross sections. 
\vskip .5cm
\centerline{\bf CONCLUSIONS AND SUGGESTIONS}
\vskip .3cm
We have seen that the VPI analysis is able to reproduce the Mainz\cite{mainz}
values for A$_{1/2}$, A$_{3/2}$, and the E2/M1 ratio (both at the resonance
energy and at the pole) by either (a) removing {\it all} pre-1980 differential
cross sections or (b) removing two sets\cite{gz74} of $\pi^0 p$ 
differential cross sections measured at Bonn in the 1970's. Statements made
by the authors of the BNL analysis\cite{bnl} imply the converse. By using
the older Bonn cross sections in their data set, they were able to obtain an
E2/M1 ratio of $-1.3$\%, a value consistent with the VPI result. There are
also preliminary indications\cite{rick} that a qualitatively similar 
effect can be seen in the RPI analysis\cite{rpi}, which is based on an 
effective Lagrangian approach.                                         
It is therefore vitally important to verify the forward and backward Bonn 
cross sections for neutral pion photoproduction. [The scale and shape 
of the recent BNL differential cross sections\cite{bnl} present yet another 
problem when compared with previous 
Bonn and Mainz cross section measurements.]  
According to Beck\cite{beckpc}, there have been wider angle measurements of
the differential cross section for $\pi^0 p$ photoproduction at Mainz. Given
the above comments, these could be crucial to a more definitive resolution 
of the E2/M1 problem. 

The discrepancy between the BNL and VPI/RPI/Mainz values for A$_{3/2}$ is
likely related to the larger BNL differential cross sections. Until these
larger cross sections are understood, every determination of the 
$\Delta$(1232) photo-decay amplitudes will require some comment 
on this issue. 
It is amusing to compare lattice predictions\cite{lattice} with 
the most recent determinations of this amplitude. 
The lattice value for A$_{3/2}$,
$-195(34)\times 10^{-3}$ GeV$^{-1/2}$, lies just outside the VPI range,
$-250(8)\times 10^{-3}$ GeV$^{-1/2}$, with nearly overlapping errors. 
The BNL value, ($-268.9\pm 2.8 \pm 4.9)\times 10^{-3}$ GeV$^{-1/2}$, lies  
somewhat further from the lattice result. 
We may see lattice results\cite{derek} with errors reduced
by a factor of 2 or 3 in the not-too-distant future. This would certainly 
make the above comparison much more interesting. 

R.W. thanks R.~Beck, R.M.~Davidson, N.C.~Mukhopadhyay, 
A.~Sandorfi, and L.~Tiator for numerous
helpful communications and for sharing their data and multipole analyses.
This work was supported in part by a U.S. Department of Energy Grant No.
DE-FG02-97ER41038.

\end{document}